\documentclass{PoS}
\usepackage{fixltx2e,amsmath}
\usepackage{amssymb}

\title{Theory overview of $B_{s,d}\to \mu^+\mu^-$ decays}

\ShortTitle{Theory overview of $B_{s,d}\to \mu^+\mu^-$ decays}

\author{\speaker{Robert Knegjens}\\
TUM Institute for Advanced Study,
Lichtenbergstr. 2a,
D-85748 Garching, Germany\\
        E-mail: \email{robert.knegjens@tum.de}}

\abstract{
In this talk I give a theoretical overview of the rare decays $B_{s} \to \mu^+ \mu^-$ and $B_{d} \to \mu^+ \mu^-$.
The branching ratios of these decays are promising probes of New Physics, both independently and relative to each other.
Recent experimental progress at the LHC has confirmed the existence of the $B_{s} \to \mu^+ \mu^-$ decay, and has not revealed any large signals of New Physics that may have been present.
This raises the question of whether moderate New Physics effects can be identified in the LHC era.
To that end I review several important developments in the Standard Model branching ratio predictions, and discuss how the latest measurements currently constrain New Physics.
Furthermore, I highlight how a time-dependent analysis of $B_{s} \to \mu^+ \mu^-$, which may be feasible at the upgraded CMS and LHCb detectors, can complement the search for and identification of New Physics.
}

\FullConference{XXII. International Workshop on Deep-Inelastic Scattering and Related Subjects,\\
		28 April - 2 May 2014\\
		Warsaw, Poland}

\def\Bqmm{{B_q\to\mu^+\mu^-}}
\def\Bsmm{{B_s\to\mu^+\mu^-}}
\def\Bdmm{{B_d\to\mu^+\mu^-}}

\def\BRsmm{{\overline{\rm BR}({B_s\to\mu^+\mu^-})}}

\begin{document}

\section{Introduction}

The decays $\Bsmm$ and $\Bdmm$ are promising probes of New Physics (NP) due to their rarity in the Standard Model (SM), clean theoretical description and experimental accessibility.
%
To elucidate these features let us consider the low-energy effective Hamiltonian describing these two transitions:
\begin{equation}
	{\cal H}_{\rm eff} = -\frac{G_F\,\alpha}{\sqrt{2}\pi}\left\{
		V_{tq}^* V_{tb}\,  \sum_{i}^{10,S,P} \left( C_i\,{\cal O}_i + C'_i\,{\cal O}'_i\right) 
		+ {\rm h.c}\right\}
		\label{mm:effOPE}
\end{equation}
with $q=\{s,d\}$, where the contributing operators are
${\cal O}_{10} = (\bar q\gamma_\mu P_L b)(\bar l \gamma^\mu \gamma_5 l)$, 
${\cal O}_{S} = (\bar q P_R b)(\bar l l),$ and
${\cal O}_{P} = (\bar q P_R b)(\bar l \gamma_5 l)$, as well as the primed versions given by the interchange $P_L\leftrightarrow P_R$.
The fully leptonic final state means that the generally troublesome hadronic matrix elements present in mesonic decays can be described by just one parameter, $\langle 0 | {\cal O}_i | \overline{B}_q^0\rangle \propto f_{B_q}$, the $B_q$ meson decay constants, 
which are most precisely calculated using lattice QCD methods.
Consequently the branching ratios of these decays, which owing to their rareness are the primary observables, can be cleanly expressed as 
\begin{align}
        {\rm BR}(B_q\to\mu^+\mu^-)
        =& {\frac{G_{\rm F}^2 \alpha^2 M_{B_q}}{16\pi^3} \sqrt{1-4\frac{m_\mu^2}{M_{B_q}^2}}}|V_{tb}V^*_{tq}|^2 |C_{10}^{\rm SM}|^2 \tau_{B_q} f_{B_q}^2 m_\mu^2 \left(|P|^2 + |S|^2 \right),
        \label{eqn:BRexpr}
\end{align}
where the terms
\begin{align}
	P &\equiv \frac{C_{10} - C'_{10}}{C_{10}^{\rm SM}} + \frac{m_{B_q}^2}{2m_\mu}\left(\frac{1}{m_b + m_q}\right) \left( \frac{C_{P} - C'_{P}}{C_{10}^{\rm SM}}\right)\equiv |P|e^{i\varphi_P}, \notag\\
	S &\equiv \sqrt{1 - \frac{4\,m_\mu^2}{m_{B_q}^2}}\frac{m_{B_q}^2}{2m_\mu}\left(\frac{1}{m_b + m_q}\right) \left( \frac{C_{S} - C'_{S}}{C_{10}^{\rm SM}}\right)\equiv |S|e^{i\varphi_S},
	\label{PSdefn}
\end{align}
are associated with the two partial wave states that the muons can be produced in.
Namely, $P$ and $S$ correspond to amplitudes with a pseudoscalar or scalar lepton density, corresponding to a CP-odd S-wave or a CP-even P-wave, respectively.
In the SM the dominant topologies responsible for the required $b\to q$ flavour-changing neutral current (FCNC) are electroweak (EW) box and penguin loops. 
Because these processes contribute only to the operator ${\cal O}_{10}$, in the SM $P=1$ and $S=0$, giving a purely CP-odd state.

The direct dependence of the branching ratios on the muon mass is due to the helicity suppression of the vector operator ${\cal O}_{10}$.
This suppression is not present for the pseudoscalar and scalar operators ${\cal O}_P$ and ${\cal O}_S$, as seen in \eqref{PSdefn}.
Therefore the $\Bqmm$ decays are especially sensitive to detecting (pseudo)scalar NP.
A popular example is the Minimal Supersymmetric Standard Model (MSSM) with large $\tan\beta$, for which the scalar and pseudoscalar Wilson coefficients scale as $\tan^3\beta$ and thereby possibly enhance the expected branching ratios by orders of magnitude~\cite{Choudhury:1998ze}.

The ratio of the branching ratios of these two decays is also an important probe of NP.
In particular in models of NP that violate the Minimal Flavour Violation (MFV) assumption, which posits that flavour symmetry is only broken by the Yukawa terms of the SM, the relation
\begin{equation}
    \frac{{\rm BR}(B_s\to \mu^+\mu^-)}{{\rm BR}(B_d\to \mu^+\mu^-)} \simeq \left|\frac{V_{ts}}{V_{td}}\right|^2 \frac{f_{B_s}^2}{f_{B_d}^2} \frac{\tau_{B_s}}{\tau_{B_d}} (\approx 32).    \label{eqn:MFVreln}
\end{equation}
will also be violated.

There has been steady experimental progress for the $B_q\to\mu^+\mu^-$ decays over the past decade.
Around the start of the LHC the Tevatron had succeeded in setting upper bounds for the branching ratios of $B_s\to \mu^+\mu^-$ and $B_d\to\mu^+\mu^-$ of $40\times 10^{-8}$ and $6\times 10^{-9}$ at a 95\% confidence level  \cite{Aaltonen:2011fi}.
Compared to the SM predictions of that time, $(3.2\pm 0.2)\times 10^{-9}$ and $(1.0 \pm 0.1)\times 10^{-10}$~\cite{Buras:2010mh}, respectively, the LHC could have witnessed an order of magnitude enhancement.
Yet four years later, experimental progress from the LHC has now culminated in the following combinations from the LHCb and CMS experiments
~\cite{Aaij:2013aka,Chatrchyan:2013bka,CMSandLHCbCollaborations:2013pla}:
\begin{align}
    \overline{\rm BR}(B_s\to \mu^+\mu^-) &= (2.9\pm 0.7)\times 10^{-9}\quad (> 5\,\sigma),\label{eqn:expCombined} \\
    \overline{\rm BR}(B_d\to \mu^+\mu^-) &= (3.6^{+1.6}_{-1.4})\times 10^{-10}\quad (< 3\,\sigma),\label{eqn:Bdmm_meas}
\end{align}
where the signal significance is also given, indicating that {\it evidence}\/ for the latter decay is still missing.
We see that the $\Bsmm$ decay is in the ballpark of its SM prediction, ruling out a smoking gun signal of NP and raising the important question of whether smallish NP in this decay can be identified in the LHC era.

\section{Progress in Standard Model branching ratio predictions}\label{sec:SMimprove}

In Refs~\cite{DeBruyn:2012wj,DescotesGenon:2011pb} it was pointed out that defining a single branching ratio for the $B_q$ meson systems -- consisting of two separate mass-eigenstates -- is in general ambiguous.
The definition used by experiments (now denoted ${\overline{\rm BR}}$) follows from counting all available events, and is thereby proportional to the time-integrated untagged decay rate.
In contrast, a branching ratio has been theoretically calculated by summing the rates of the two flavour states at $t=0$ and normalizing with the mean $B_q$ lifetime $\tau_{B_q}$ (e.g.\ see \eqref{eqn:BRexpr}).
The two definitions only agree in the limit of no lifetime difference for the two mass-eigenstates i.e. $y_q = (\tau_{q,\rm H} - \tau_{q,\rm L})/(\tau_{q,\rm H} + \tau_{q,\rm L}) \to 0$.
Though this is almost true for the $B_d$ system, the $B_s$ system has a sizable difference:  $y_s = 0.075\pm 0.012$~\cite{Aaij:2013oba}.
This necessitates a dictionary to convert between the two definitions, given by~\cite{DeBruyn:2012wj}
\begin{equation}
    \overline{\rm BR}(B_q \to \mu^+\mu^-) =
    {\rm BR}(B_q \to \mu^+\mu^-)\left[\frac{1 + y_q\,{{\cal A}_{\Delta\Gamma}(B_q \to \mu^+\mu^-)}}{1-y_q^2}\right]
\end{equation}
for the decays at hand, where
\begin{equation}
    {{\cal A}_{\Delta\Gamma}(B_q \to \mu^+\mu^-)} =
\frac{{\Gamma(B_{q,{\rm H}} \to \mu^+\mu^-)} - {\Gamma(B_{q,{\rm L}} \to \mu^+\mu^-)}}{{\Gamma(B_{q,{\rm H}} \to \mu^+\mu^-)} + {\Gamma(B_{q,{\rm L}} \to \mu^+\mu^-)}}
\label{eqn:ADG}
\end{equation}
is the {\it mass-eigenstate rate asymmetry}.
As already discussed, the final state muons in the SM are produced in a CP-odd configuration.
Because the $B_s^0$--$\overline{B}_s^0$ mixing phase cancels against the CP violating phase appearing in the decay mode, this transition proceeds purely via the heavy mass-eigenstate\footnote{CP violating phases from internal charm quarks in the decay loops do not cancel but are dynamically suppressed and thereby negligible.}.
Thus ${{\cal A}_{\Delta\Gamma}(B_q \to \mu^+\mu^-)}=1$ in the SM, leading to a maximal correction of 8\% for the theoretical SM prediction~\cite{DeBruyn:2012wk}.

\begin{figure}
\begin{center}
\includegraphics[height=2.95cm]{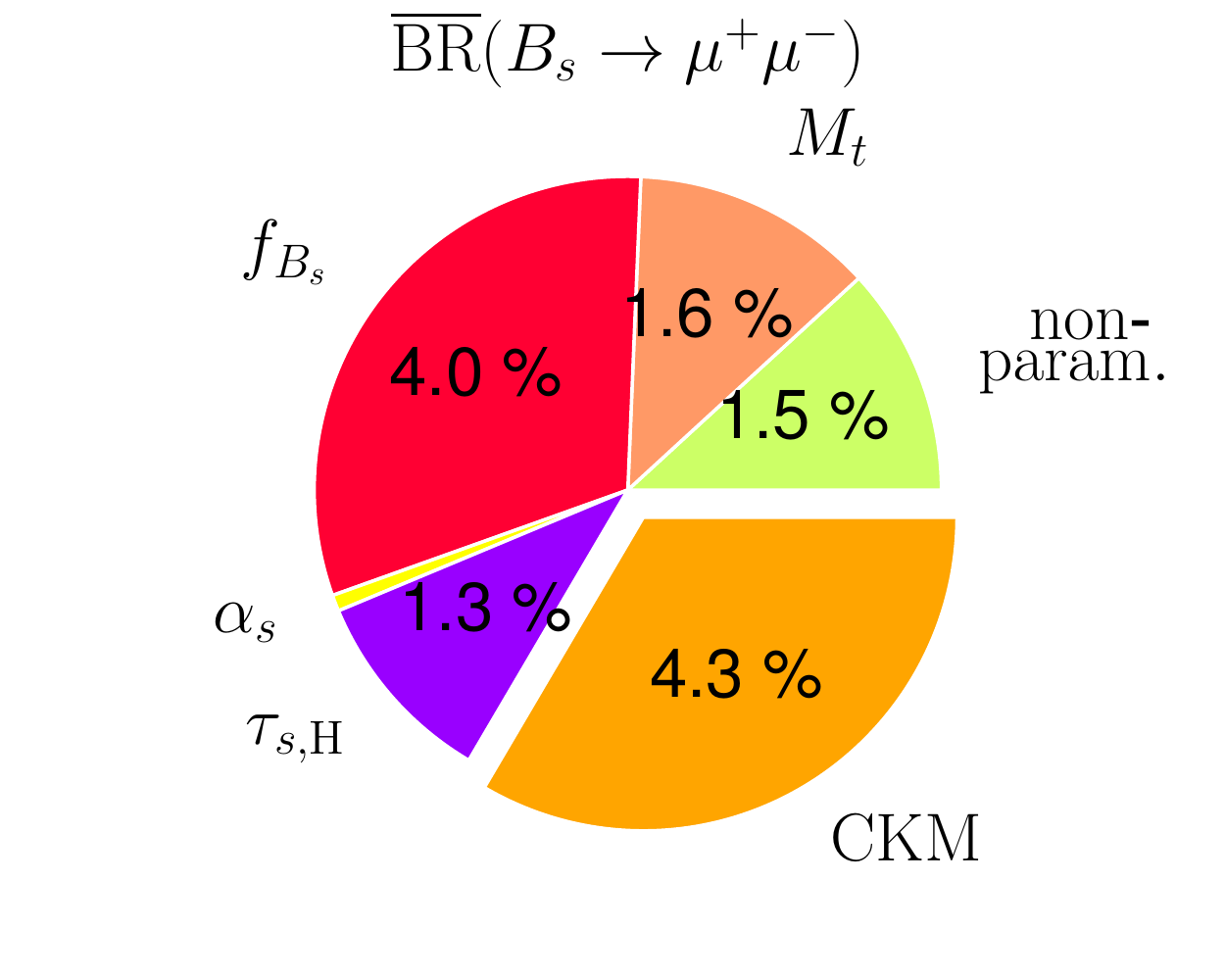}
\includegraphics[height=2.95cm]{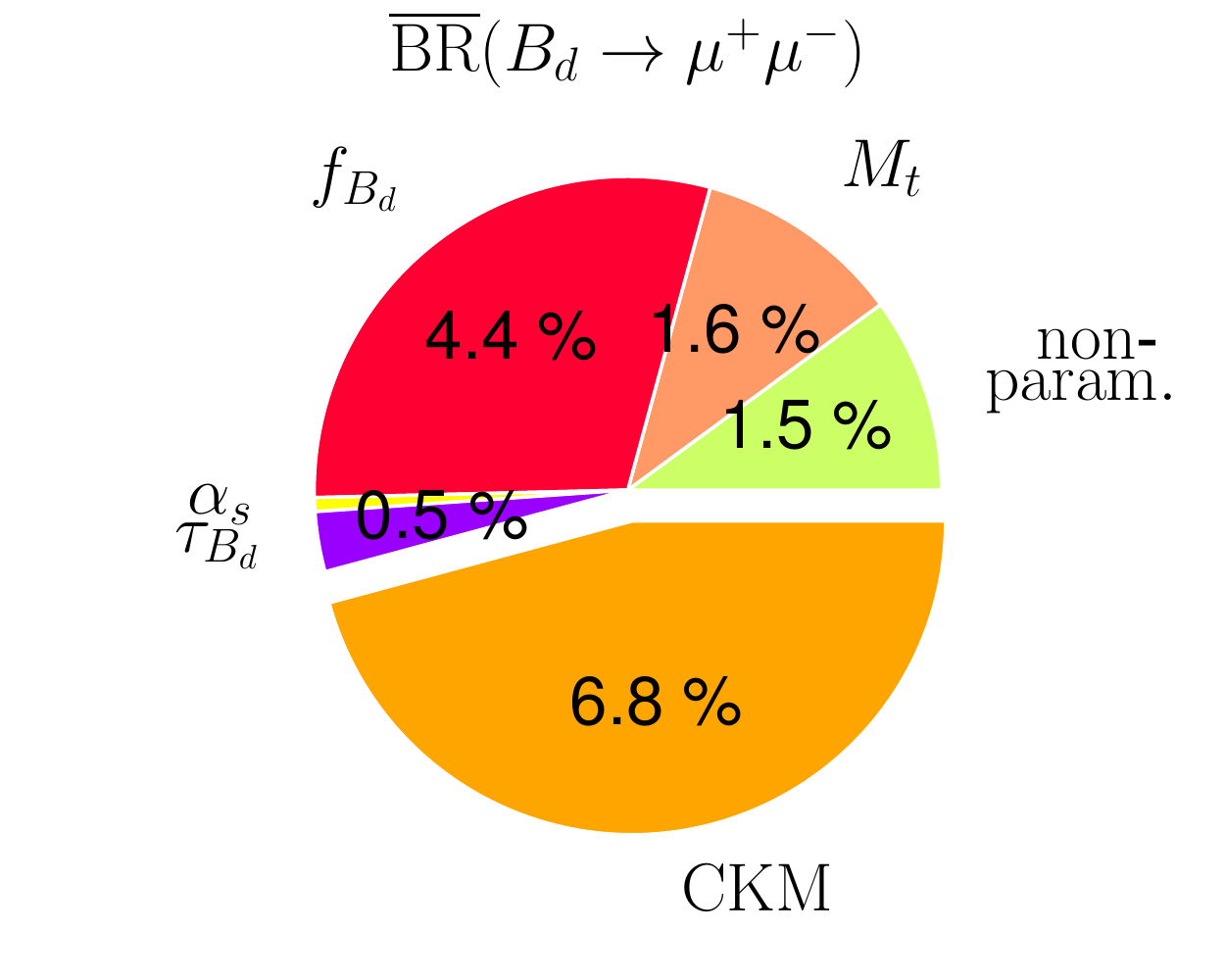}
\includegraphics[height=2.95cm]{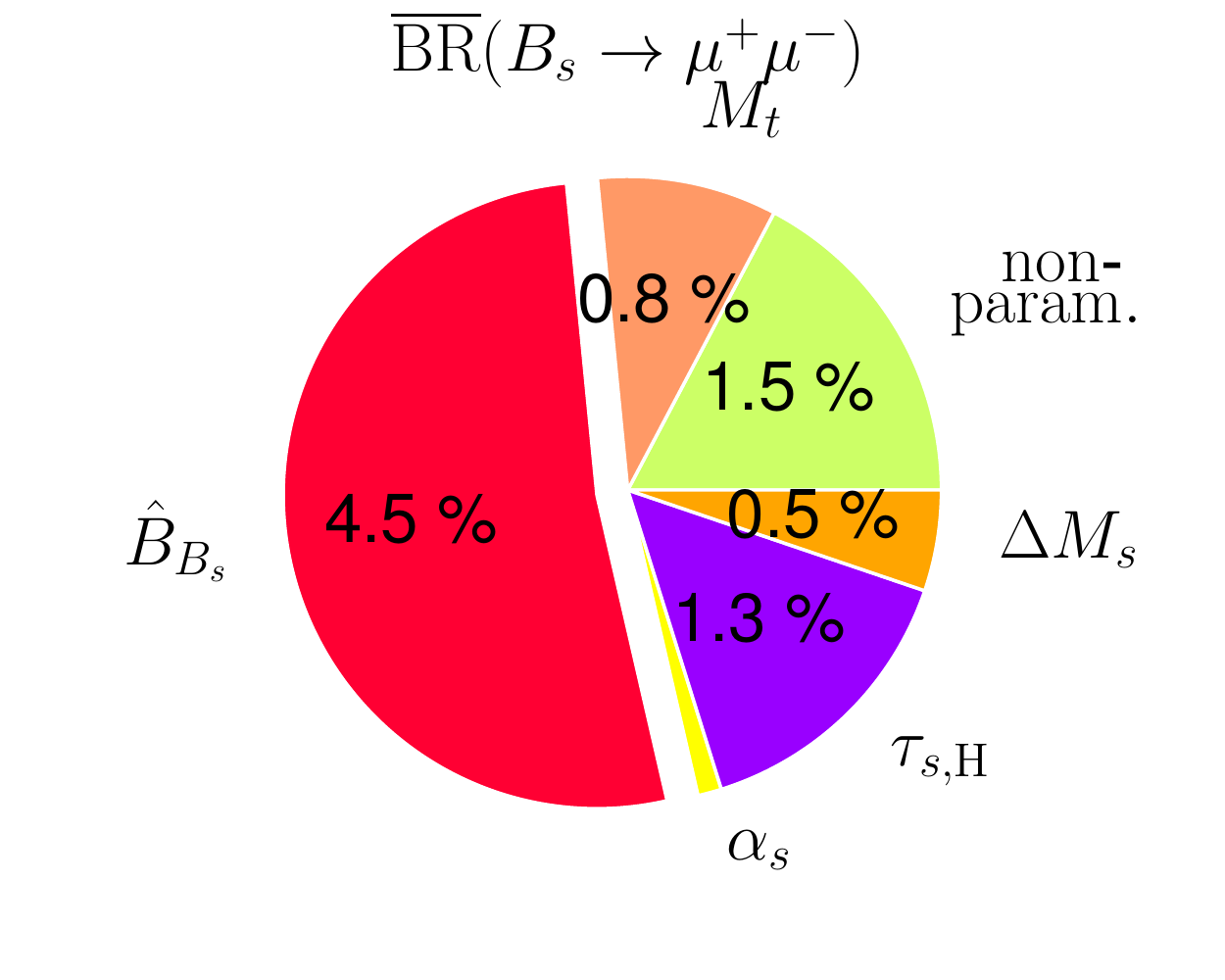}
\includegraphics[height=2.95cm]{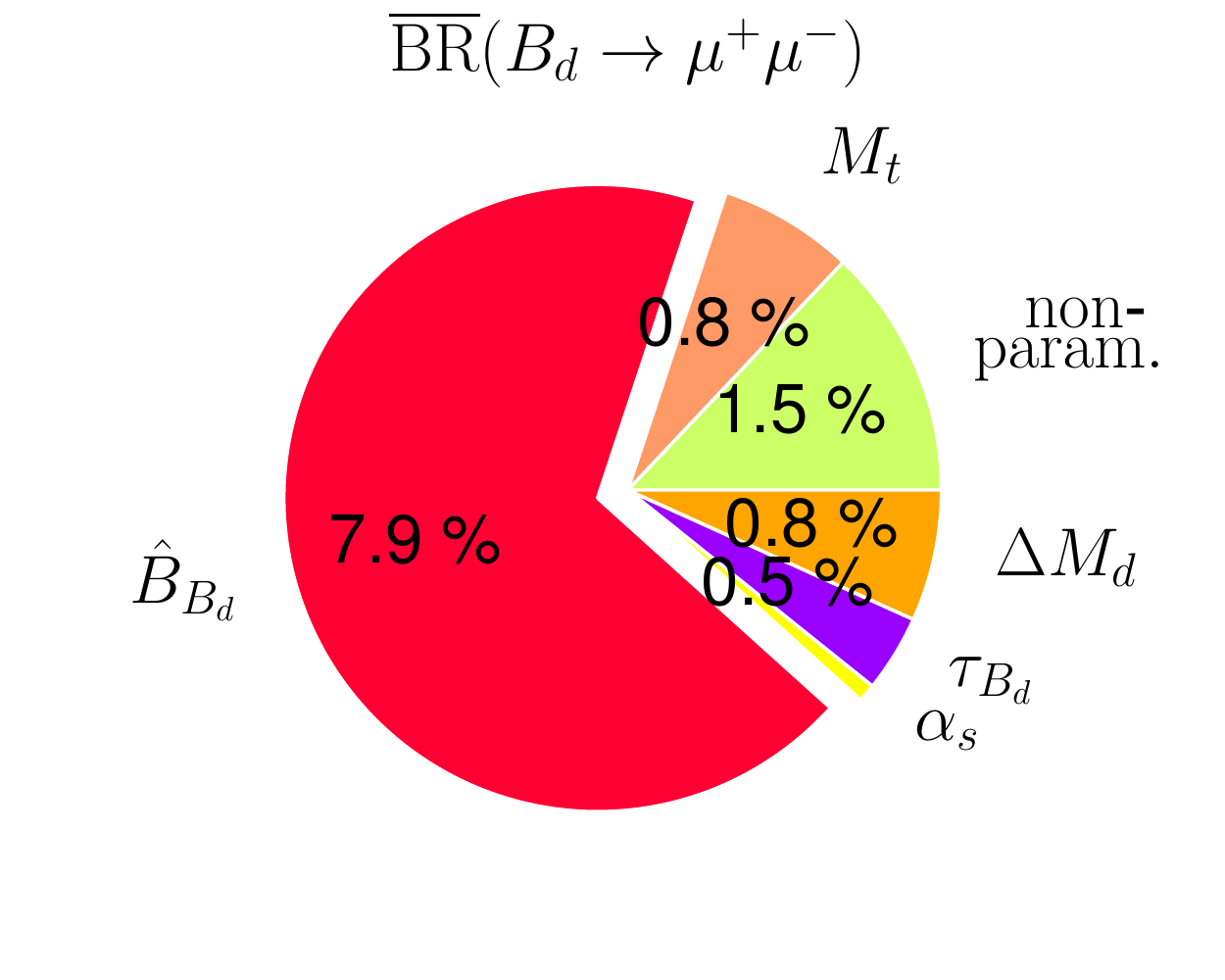}
\end{center}
\caption{Error budgets for the Standard Model predictions of $\overline{\rm BR}(B_s\to\mu^+\mu^-)$ and $\overline{\rm BR}(B_d\to\mu^+\mu^-)$ given in \eqref{eqn:SMpreds} and \eqref{eqn:SMaltpreds}. The two budgets on the right have the dependence on CKM parameters and $f_{B_q}$ substituted for $\Delta M_q$ and $\hat{B}_{B_q}$ as proposed in Ref.~\cite{Buras:2003td}. The relative errors summed in quadrature are 6.4\%, 8.5\%, 5.1\% and 8.8\%, respectively.}
\label{fig:pieplots}
\end{figure}

There has been impressive recent progress in the calculation of perturbative loop corrections for the $\Bqmm$ processes in the SM.
In Ref.~\cite{Bobeth:2013tba} the next-to-leading order EW corrections have been calculated, which reduce a 7\% scale and scheme uncertainty to less than 1\%.
And in Ref.~\cite{Hermann:2013kca} the next-to-next-to-leading order QCD corrections were calculated, reducing a scale uncertainty of 1.8\% to less than 0.2\%\footnote{
See the talk of Mikolaj Misiak in these proceedings for further details.}.
Combined with improved lattice determinations of the $B_q$ decay constants~\cite{Aoki:2013ldr}, the state of the art SM predictions are~\cite{Bobeth:2013uxa}
\begin{align}
{\overline{\rm BR}}(B_s\to\mu^+\mu^-)_{\rm SM} &= {(3.65\pm 0.23)\times 10^{-9}},\\
{\overline{\rm BR}}(B_d\to\mu^+\mu^-)_{\rm SM} &= {(1.06\pm 0.09)\times 10^{-10}}.
\label{eqn:SMpreds}
\end{align}
In the left two plots of Figure~\ref{fig:pieplots} we present the corresponding updated error budgets.
We observe that the CKM elements now give the dominant uncertainty.
Alternatively we can substitute the two dominant uncertainties with the experimental determination of the $B_q$ mass difference $\Delta M_q$ and the lattice determination of the bag parameter $\hat{B}_{B_q}$, as proposed in Ref.~\cite{Buras:2003td}.
Using inputs from Refs~\cite{Aoki:2013ldr,Beringer:1900zz}, the predictions in this case are
\begin{align}
    {\overline{\rm BR}}(B_s\to\mu^+\mu^-)_{{\rm SM}(\Delta M_s,\hat{B}_{B_s})} &= (3.53\pm 0.18)\times 10^{-9}, \\   
    {\overline{\rm BR}}(B_d\to\mu^+\mu^-)_{{\rm SM}(\Delta M_d,\hat{B}_{B_d})} &= (1.00\pm 0.09)\times 10^{-10}.
    \label{eqn:SMaltpreds}
\end{align}
In the right two plots of Figure~\ref{fig:pieplots} we present the corresponding error budgets.
The total errors are comparable in size while the central values are slightly lower.

\section{Constraining New Physics with $\boldsymbol{B_s\to\mu^+\mu^-}$}\label{sec:constraining}

Were the $\Bdmm$ central value in \eqref{eqn:Bdmm_meas} to hold in future measurements, it would be a telling signal of NP.
Unfortunately, a statistically significant {\it discovery} of this decay is only expected after the LHC upgrade~\cite{CMS:2013vfa}.
Furthermore, only by the end of the LHCb and CMS upgrade runs is the ratio given in \eqref{eqn:MFVreln} expected to be measured with a precision of 35\% and 21\%, respectively~\cite{Bediaga:2012py,CMS:2013vfa} (see Ref.~\cite{Bobeth:2014tza} for a summary).
We therefore proceed to consider how the discovery of $\BRsmm$, which is currently $1\,\sigma$ below its SM prediction, constrains possible NP contributing to its effective operators.

The coefficients $C_{10}^{(\prime)}$ can receive NP contributions from new vector-like particle interactions.
Examples include models with $Z'$ gauge bosons~\cite{Buras:2012jb}, modified $Z$ couplings~\cite{Guadagnoli:2013mru,Straub:2013zca}, or vector-like fermions~\cite{Buras:2013td}.
Due to the helicity suppression of these vector-like operators, constraints from other $b\to s\ell^+ \ell^-$ transitions also compete, in particular $B_d \to K^*\mu^+ \mu^-$~\cite{Altmannshofer:2012ir}.
A good approach is therefore to make a global analysis of $b\to s(\ell^+ \ell^-,\gamma)$ transitions, as recently performed in Ref.~\cite{Beaujean:2013soa}\footnote{See the talk of Danny van Dyk in these proceedings}.

\begin{figure}
\begin{center}
\includegraphics[width=5.0cm]{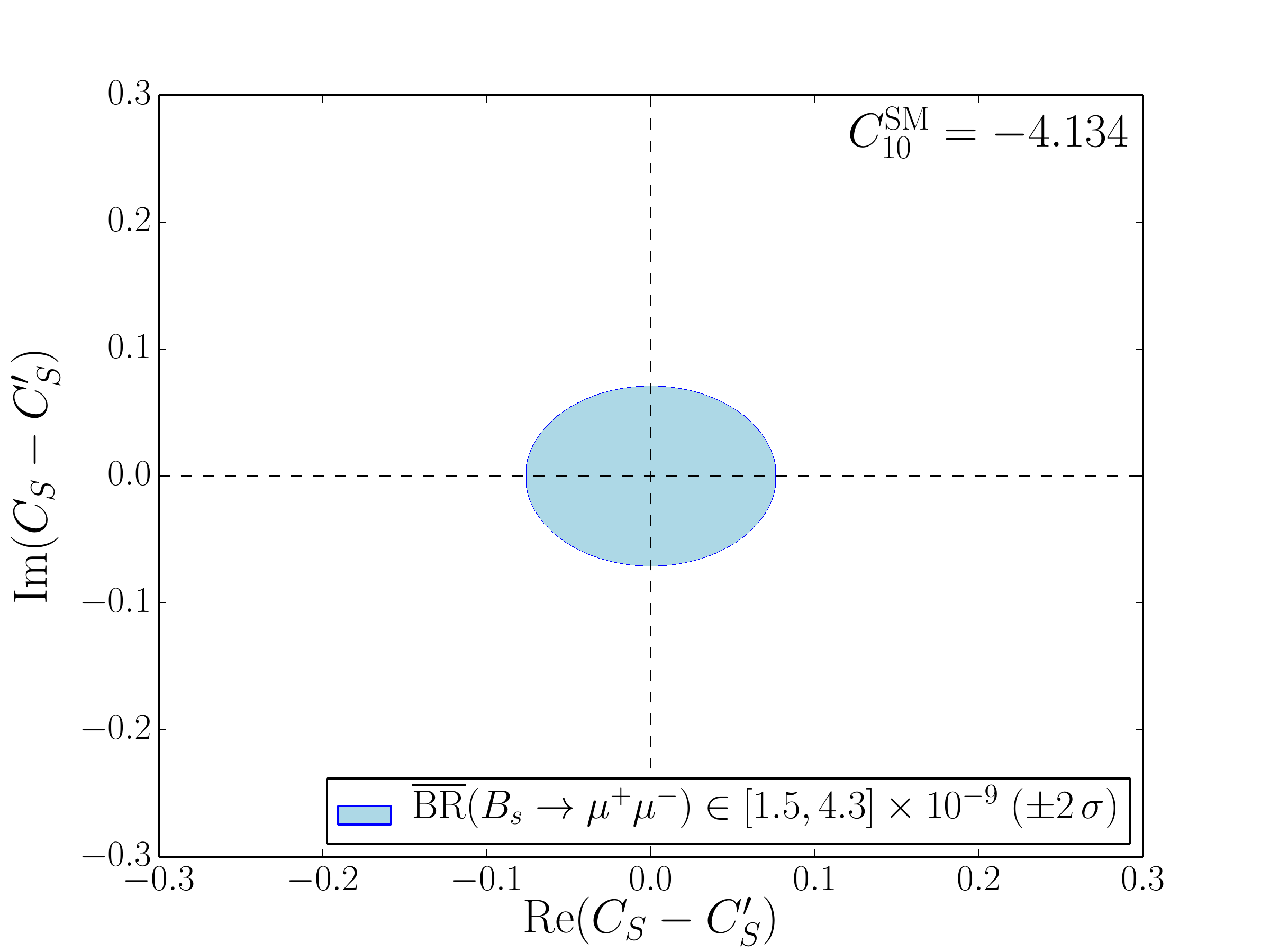}
\includegraphics[width=5.0cm]{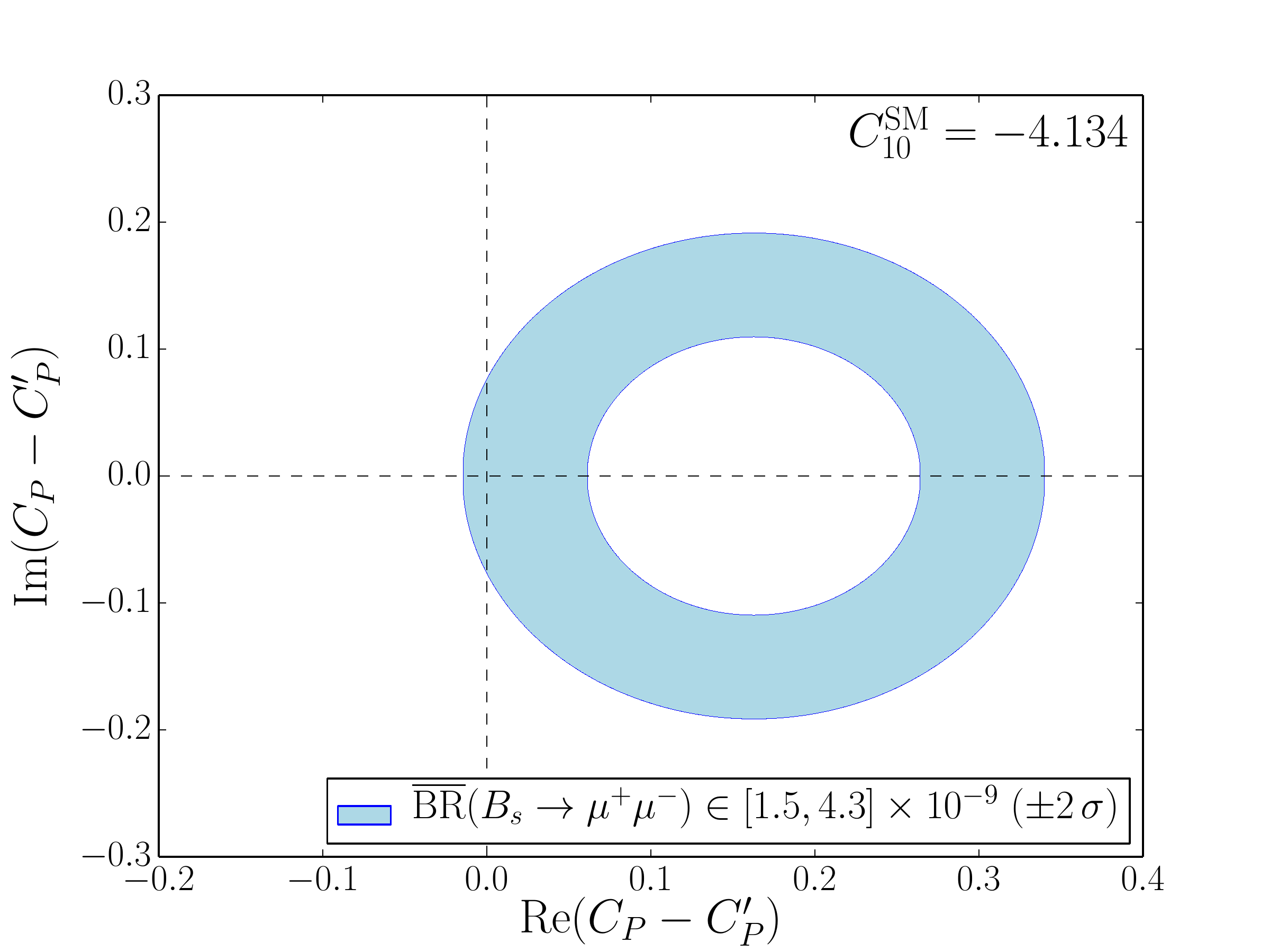}
\end{center}
\caption{Bounds from the combined measurement of $\overline{\rm BR}(B_s\to\mu^+\mu^-)$ in \eqref{eqn:expCombined} on the scalar and pseudoscalar Wilson coefficient combinations $C_S -C'_S$ (left panel) and $C_P -C'_P$ (right panel).}
\label{fig:scalar_bounds}
\end{figure}

$\Bsmm$ has the best sensitivity for the (pseudo)scalar Wilson coefficient combinations $C_S-C'_S$ and $C_P-C'_P$, which enter without helicity suppression.
These coefficients are affected by models with flavour-changing (pseudo)scalar particles~\cite{Buras:2013rqa}, in particular two Higgs doublet models (2HDM), including the MSSM.
In Figure~\ref{fig:scalar_bounds} we show the regions still allowed within $2\,\sigma$ interval of this branching ratio.
The $C_S-C'_S$ are more tightly constrained compared to $C_P-C'_P$ because the $C^{(\prime)}_S$ do not interfere with the SM and can thereby only increase $\BRsmm$ above its SM prediction.
Although not yet competitive in experimental precision, complementary constraints for the combinations $C_S + C'_S$ and $C_P + C'_P$ are given by the decay $B_d\to K\mu^+\mu^-$\cite{Becirevic:2012fy}.
In the MSSM the strong dependence on $\tan\beta$ when this parameter is large leads to strong constraints when the dominant FCNC loops interfere constructively with the SM~\cite{Altmannshofer:2012ks}.
In the case of destructive interference $\Bsmm$ is much less constraining, which is also the situation for non-supersymmetric aligned 2HDMs in general~\cite{Li:2014fea}.

%

%
%

\section{$\Bsmm$ time-dependent observables and outlook}\label{sec:mestates}

Because in the SM the $\Bsmm$ transition proceeds purely via the heavy mass-eigenstate, a contribution from the light mass-eigenstate would be a clear signal of NP.
Specifically, its presence would signal new CP violating phases or the presence of scalar operators:
\begin{equation}
    \Gamma{(B_{s,{\rm L}}}\to \mu^+\mu^-) \propto |{P}|^2\underbrace{\sin^2({\varphi_P} - \phi_s^{\rm NP}/2)}_{\rm new\ CP\ phases} + \underbrace{|{S}|^2\cos^2({\varphi_S} - \phi_s^{\rm NP}/2)}_{\rm scalar\ operators}.
\end{equation}
The relevant observable for probing this NP is the mass-eigenstate rate asymmetry ${\cal A}_{\Delta\Gamma}^{\mu\mu}$ defined in \eqref{eqn:ADG}, which depends on $P$ and $S$ as~\cite{DeBruyn:2012wk}
\begin{equation}
    {\cal A}_{\Delta\Gamma}^{\mu\mu} = 
    \frac{|{P}|^2\cos\left(2{\varphi_P} {- \phi_s^{\rm NP}}\right) - |{S}|^2\cos\left(2{\varphi_S} {- \phi_s^{\rm NP}}\right)}{|{P}|^2 + |{S}|^2}.
\end{equation}
It can be extracted directly from a fit to the time-dependent untagged $\Bsmm$ rate, or from an effective lifetime measurement.

Due to the small $B_s$ lifetime difference $y_s$, in order to achieve an uncertainty of 10\% for ${\cal A}_{\Delta\Gamma}^{\mu\mu}$ a precision of 1\% for the effective lifetime measurement is needed. 
This will be challenging for the LHCb experiment, for which a better than 5\% uncertainty on the effective lifetime is projected~\cite{DeBruyn:2012wk} with the $\sim 350$ events expected by the end of the upgrade run.
In comparison, the CMS experiment expects 2096 events by the end of its high luminosity upgrade~\cite{CMS:2013vfa}.
Both experiments project a similar ultimate precision on the $\Bsmm$ branching ratio of $\approx 10\%$ due to the dominance of systematic errors, specifically the ratio of fragmentation functions $f_s/f_d$ necessary for normalization.
However, as the asymmetry ${\cal A}_{\Delta\Gamma}^{\mu\mu}$ does not depend on this dominant systematic error, the desired precision might be reachable with the larger CMS data sample.

\begin{figure}
\begin{center}
\includegraphics[width=4.59cm]{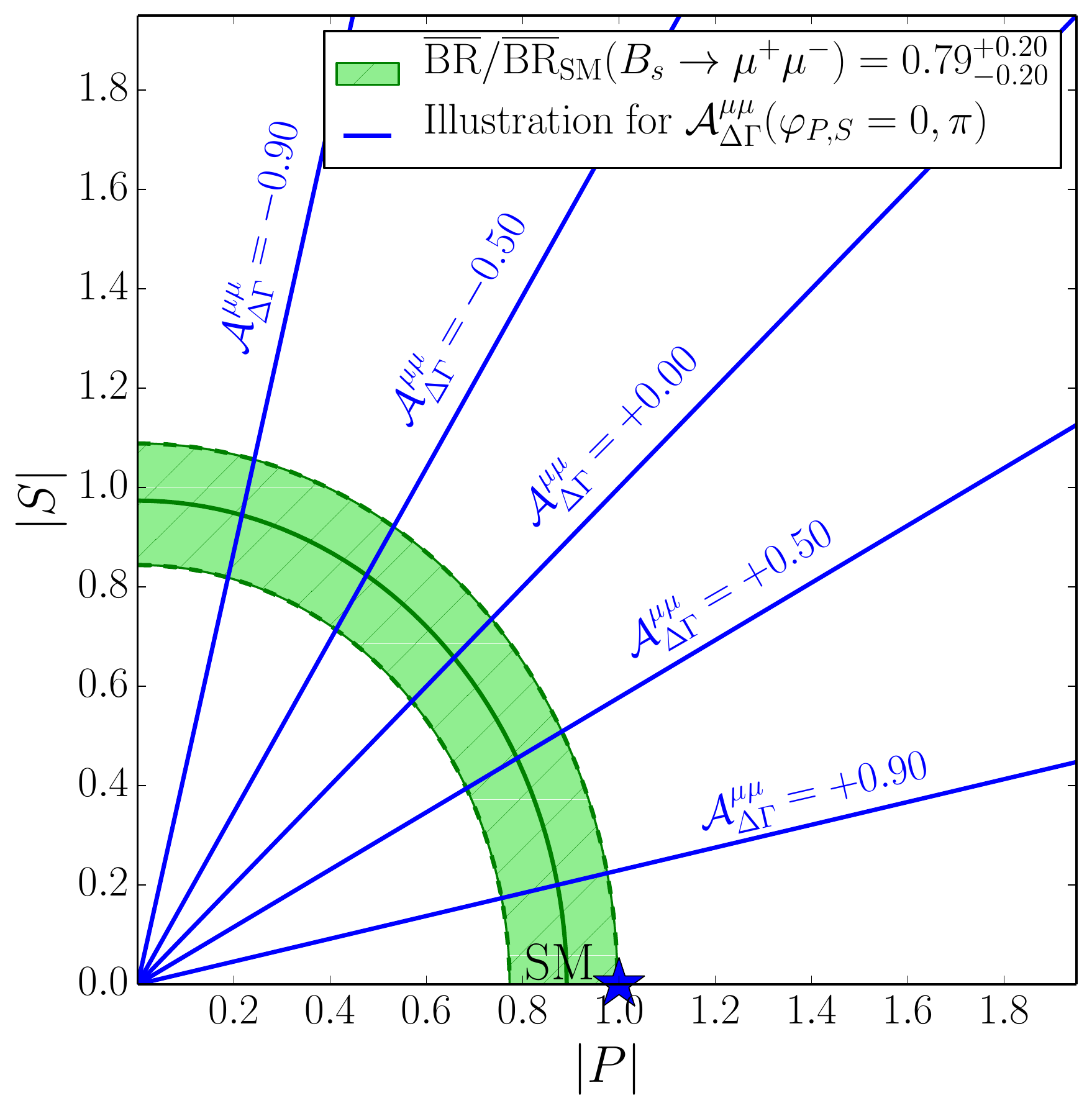}
\includegraphics[width=6.0cm]{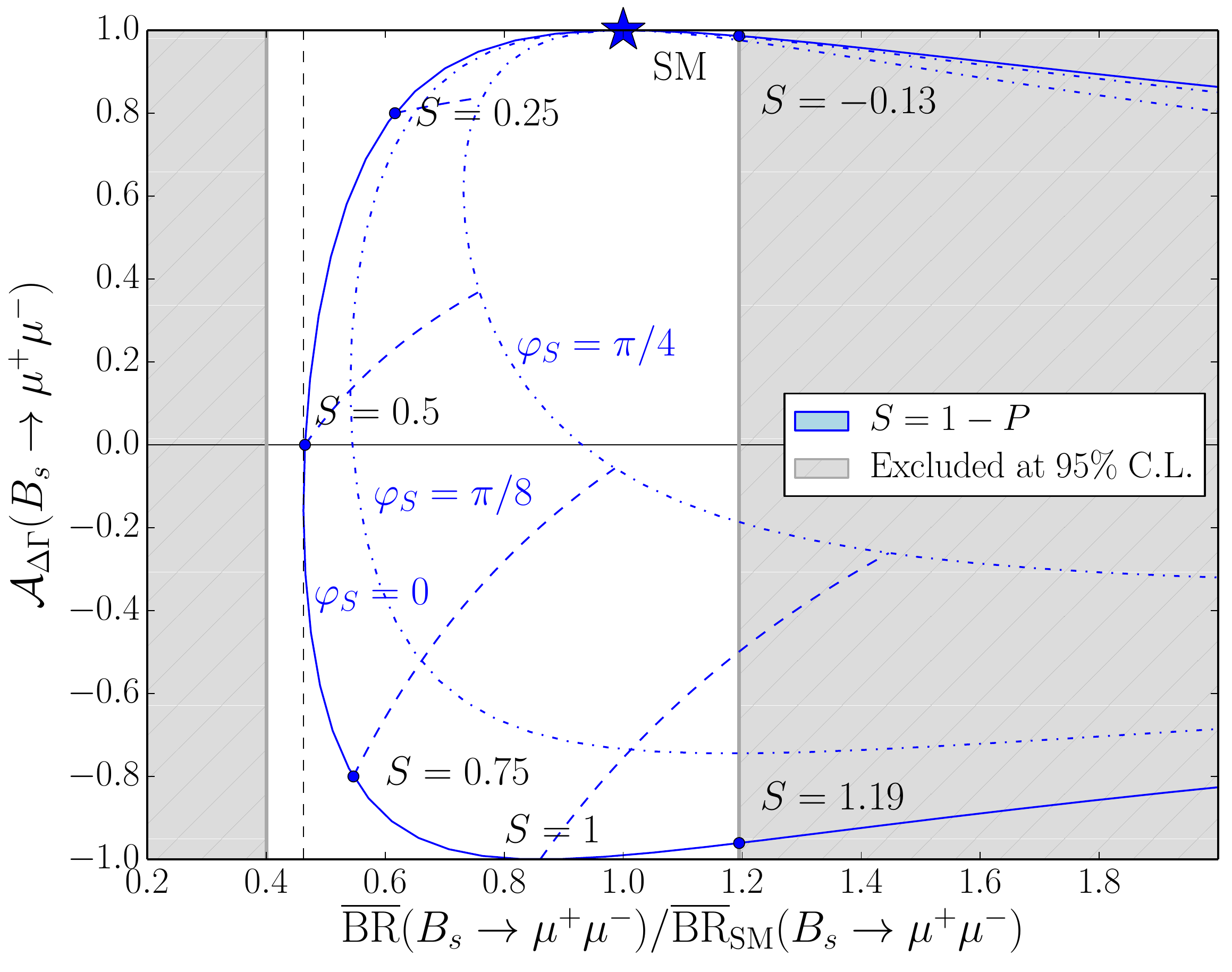}
\end{center}
\caption{{\it Left panel:} relation between $|S|$ and $|P|$ for varying values of $\BRsmm$ and ${\cal A}_{\Delta\Gamma}^{\mu\mu}$ in a NP scenario with trivial new CP violating phases (updated from Ref.~\cite{DeBruyn:2012wk}). {\it Right panel:} relations between $\BRsmm$ and ${\cal A}_{\Delta\Gamma}^{\mu\mu}$ in a NP scenario where $S=1-P$, such as a decoupled 2HDM (updated from Ref.~\cite{Buras:2013uqa}).}
\label{fig:ADGplots}
\end{figure}

The observables $\BRsmm$ and ${\cal A}_{\Delta\Gamma}^{\mu\mu}$ complement each other in identifying and discriminating NP.
In the absence of new CP violating phases, for example, $\BRsmm$ probes the sum of $|P|^2$ and $|S|^2$ whereas ${\cal A}_{\Delta\Gamma}^{\mu\mu}$ the difference, as shown in the left panel of Figure~\ref{fig:ADGplots}.
An interesting scenario of NP in which the full range of ${\cal A}_{\Delta\Gamma}^{\mu\mu}$ is possible without new CP violating phases is when scalar and pseudoscalar NP contributions are on the same footing i.e. $C^{(\prime)}_S \simeq \pm C^{(\prime)}_P$ or $S\simeq \pm (1-P)$~\cite{Buras:2013uqa}.
This is realised for example in decoupled 2HDMs, such as in the MSSM at large $\tan\beta$.
We show this scenario in the right panel of Figure~\ref{fig:ADGplots}, where also the lower bound $\BRsmm/\BRsmm_{\rm SM} \geq (1-y_s)/2$ present in such models is visible.
We note that even if the branching ratio measurement settles exactly on the SM value, a measurement of the sign of ${\cal A}_{\Delta\Gamma}^{\mu\mu}$ is needed to confirm or exclude a hidden NP solution.

The discovery of $B_s\to\mu^+\mu^-$ at the LHC is a milestone for the $B_q\to\mu^+\mu^-$ decays, providing strong constraints on NP and prompting improvements in the SM predictions.
We hope that the LHC will proceed to deliver a discovery of $B_d\to\mu^+\mu^-$ as well as a decay-time profile of $B_s\to\mu^+\mu^-$, both of which would provide vital complementary information on the presence of NP.


\acknowledgments

I would like to thank Martin Gorbahn for the invitation to this workshop, and the organizers for their great organization and hospitality.
I would also like to thank Christoph Bobeth for useful discussions related to these proceedings and his comments on this manuscript.
This work was supported by the ERC Advanced Grant project ``FLAVOUR'' (267104).

\bibliographystyle{h-physrev}
\bibliography{references}

\end{document}